# A SCADE Model Verification Method Based on B-Model Transformation

Xili Hou[1], Keming Wang[2+,3], Huibing Zhao[1], Ruiyin Shi[1]

1. Beijing Jiaoda Signal Technology Co., Ltd., Beijing 102206, China
2. School of Computing and Artificial Intelligence, Southwest Jiaotong University, Chengdu 611756, China
3. National-Local Joint Engineering Laboratory of System Credibility Automatic Verification, Southwest Jiaotong University, Chengdu 611756, China

E-mail addresses: b.method.xili.hou@gmail.com (X.L. Hou), kmwang@swjtu.edu.cn (K.M. Wang), zhaohuibing@jd-signal.com (H.B. Zhao), shiruiyin@jd-signal.com (R.Y. Shi)

**Abstract:** Due to the limitations of SCADE models in expressing and verifying abstract specifications in safety-critical systems, this study proposes a formal verification framework based on the B-Method. By establishing a semantic equivalence transformation mechanism from SCADE models to B models, a hierarchical mapping rule set is constructed, covering type systems, control flow structures, and state machines. This effectively addresses key technical challenges such as loop-equivalent transformation proof for high-order operators and modeling of temporal logic storage structures. The proposed method innovatively leverages the B-Method's abstraction capabilities in set theory and first-order logic, overcoming the constraints of SCADE's native verification tools in complex specification descriptions. It successfully verifies abstract specifications that are difficult to model directly in SCADE. Experimental results show that the transformed B models achieve a higher defect detection rate and improved verification efficiency in the ProB verification environment compared to SCADE's native verifier, significantly enhancing the formal verification capability of safety-critical systems. This study provides a cross-model verification paradigm for embedded control systems in avionics, rail transportation, and other domains, demonstrating substantial engineering application value.

## 1. Introduction

In embedded systems, Ansys SCADE Suite is a specialized modeling tool for developing safety-critical software, widely used in aviation, nuclear power, transportation, and automotive industries. SCADE[1] (*Safety Critical Application Development Environment*), as a mainstream modeling tool in safety-critical domains, is based on the synchronous dataflow language Lustre, supporting graphical modeling and formal verification for the development of embedded automatic control systems. It has been successfully applied to avionics systems, rail transit control[2], and other fields. However, since SCADE mainly uses dataflow diagrams and state machines to describe system logic, it has certain limitations in expressing high-abstraction-level system specifications. In particular, some complex abstract specifications or abstract logic cannot be directly represented in SCADE.

The B-Method[3], as a formal method based on set theory, relational algebra, and first-order predicate logic, allows gradual refinement from abstract specifications to executable code. It supports full-cycle formal verification from high-level specifications to executable code and has accumulated a rich industrial toolchain (such as Atelier-B and ProB) along with successful case applications[4].

The B-Method excels in precise abstraction capabilities, enabling the description of high-level specifications that SCADE struggles to model directly. By transforming SCADE models into B-Method models, the strong abstraction capabilities of the B-Method can compensate for SCADE's shortcomings at the abstract level. Additionally, formal verification tools in the B-Method can be utilized to verify the correctness of SCADE models, ensuring compliance with design requirements. This integration not only enhances the formal verification capabilities of the system but also helps identify potential issues within SCADE models, thereby improving system reliability and safety. Therefore, studying how to convert SCADE models into B-Method models and perform formal verification using B-Method tools is of great theoretical and practical significance.

To address these challenges, this study proposes a transformation method from SCADE models to B-Method models, aiming to overcome the limitations of SCADE's native verification tools in abstract specification expression. By establishing formal mapping rules for type systems, control flow structures, and state machines, this method enables semantically equivalent transformation from SCADE models to B models. The proposed approach innovatively addresses key technical issues such as loop-equivalence transformation proof for high-order operators and the modeling of temporal logic storage structures. Experimental results demonstrate that the transformed B models can effectively verify abstract specifications that SCADE models fail to express, thereby enhancing system development reliability. The SCADE project files and B-Model files related to the experiment are available on GitHub: https://github.com/Alex-Hou-2024.

**2. Preliminaries**

2.1 Introduction of SCADE

SCADE is a high-level language and environment for developing safety-critical embedded control software. It has been used for more than twenty years in various application domains like avionics, nuclear plants, transportation, and automotive. SCADE has been founded on the synchronous data-flow language Lustre[5] invented by Caspi and Halbwachs. All Scade language elements in detail can be found at [11]. The SCADE language contains the following sections:

2.1.1 Type

SCADE supports multiple data types, including integers, floating-point numbers, booleans, and arrays. The type system ensures data correctness during operations while supporting polymorphic types, allowing the use of generic type variables in node definitions. This enables nodes to be applied to various data structures and operations.

2.1.2 Constant

Constants in SCADE are defined as part of the global flow and can be accessed within nodes. The values of constants remain unchanged throughout the computation process and are based on the local clock of the node. This definition makes constants important fixed reference values during model construction, suitable for fixed parameters or reference data in a system.

2.1.3 Equation

Equations are the core of the SCADE language, used to define data flow relationships within nodes. Similar to function definitions, equations describe system logic through input-output mappings. Each equation is computed synchronously within each cycle, ensuring deterministic system outputs. This determinism is a key advantage of SCADE in real-time and safety-critical systems.

2.1.4 Condition Expression

Condition expressions implement control logic, allowing the activation or deactivation of specific operations within the current cycle based on boolean values. SCADE supports *if-then-else* structures for defining conditional expressions. Additionally, enumerated condition expressions can be used to activate specific logic branches through the *activate when* structure, executing different process flows based on different enumeration values.

2.1.5 Sequential Operator

Sequential operators define sequential logic, ensuring operations execute in a predetermined order. Sequential operators are particularly critical for defining models that require memory of the previous cycle's state, playing an important role in state machines and temporal logic implementations.

2.1.6 State Machine

State machines are an essential tool for control flow in SCADE. A state machine consists of multiple states, with each state defining its own set of equations to control outputs. State transitions are triggered by boolean conditions or signals. State machines simplify complex control logic, such as mode switching and conditional activation. SCADE supports composite state machines and complex state transition structures to meet various control requirements and high-level mode management needs.

2.1.7 Higher-Order Operator

SCADE provides higher-order operators, such as *map* and *fold*, for iterative computations on array elements. These operators enable custom functions to be applied to arrays, *map* applies an operation to each element in parallel, *fold* accumulates results sequentially. These higher-order operators offer a concise and structured approach to complex array operations, making them suitable for traversal and aggregation scenarios.

2.2 Introduction of B method

The B method[3] is a formal method for software development based on set theory and first-order predicate logic. The basic component of the B method is the abstract machine. Abstract machines are divided into three levels: the *MACHINEs*, which describe the highest level of specification; the *REFINEMENTs*, which include all the intermediary steps between the specification and the code. The *REFINEMENTs* provide a way to construct stronger invariants and also to enrich a model in a step-by-step approach; The final *REFINEMENTs* are *IMPLEMENTATIONs*, which corresponds to a stage of development leading to the production of code when the language of substitutions is restricted to the B0 language, B0 is a subset of the language of substitutions and translation to C, C++, or ADA is possible in tools[3][6][7].

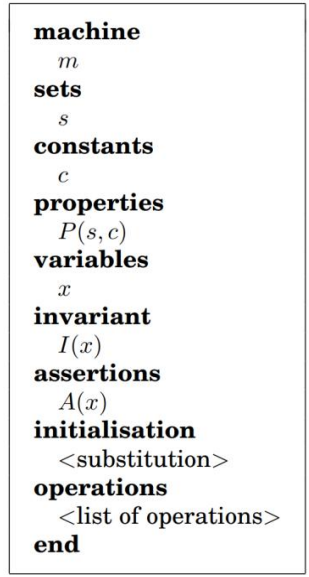

Figure 1. The structure of the abstract machine

The B method defines the Abstract Machine Notation to describe the abovementioned three levels of abstraction[3]. Figure 1 shows the structure of *machine m,* which has the declarative part to define the data, the state, and the executive part defining operations. The *refinement* and *implementation* follow the same model of *machine*[3][6][7].

In the declarative part of the model, the clause *sets* contain definitions of sets; the clause *constants* are either scalar constants of a set, or total functions from a set (or a cartesian product of sets) to a set, or else subsets of a set. The clause *properties* contain the effective definitions of *sets* and *constants*, which should always be verified by the state of the machine. The *variables* define the dynamic aspects of state variables and properties. The *invariant I(x)* defines the type of the variable *x,* which is assumed to be initialized concerning the initial conditions and it is supposed to be preserved by operations (or transitions) of the list of operations. The *invariant I(x)* is an essential feature of abstract *machine m*, *refinement*, and *implementation*. Invariants can express properties of the formal model that hold in every reachable state of the components[8], and the *invariants* should always be confirmed when the variables in *invariants* change their values.

The clause *assertion A(x)* contains the list of theorems discharged by the proof engine. The composition clauses like *sees, includes, extends, promotes, uses,* and *imports*, which are not mentioned in Figure 1, can help describe the various links between abstract machines. For example, the *includes* primitive can be used in an abstract machine or a refinement; the included component allows the including component to modify included variables by included operations; The *imports* clause of an implementation contains the list of the (possibly renamed) imported machines[3][6][7].

The executive part contains the *initialisation* and the *operations* of the abstract machine. The feasibility of the initialisation event requires that at least one value exists for the predicate defining the initial conditions; the operations should provide the set of behavior enabling the state of the machine to evolve. The proof guarantees that the B-model is coherent and respects the properties introduced in precondition, invariant, and postcondition[7]. In the first part of the model, proof obligations are generated from the mathematical theory defined. In the second part of the model, proof obligations are generated for the preservation (when calling the operation) of the invariant, and proof obligations state the correctness of safety properties with respect to the invariant[6]. More details of the B method can be read in the book of B-Book[3].

The B method is supported by the robust, commercially available tool Atelier-B[9] for project management, static checking, proof obligation generation, automatic and interactive proof, and code generation. The animator and model checker ProB[10] completes Atelier-B. B has been used in major safety-critical system applications. A summary of 25 years of development and industrial use of Atelier-B can be found in [4].

**3. Transformation Rules**

In this section, we present the overall translation approach from SCADE to B models, covering the translation of various components. This includes translating data types, constants, equations, control flows, and state machines to ensure a consistent semantic representation in the B model. The translation also addresses the handling of sequential operators and higher-order operations to maintain the functional equivalence between the SCADE and B models.

When translating from SCADE to B-Model, it is essential to ensure that the semantics and functionality are preserved. We show here how to translate SCADE into a B model. Only the mapping relationship between the most basic SCADE model and the corresponding B model is shown here. Complex SCADE models can generate corresponding B models after iteration based on these basic mapping relationships.

3.1 Type

Type declarations are used to build new user data types. Base types such as uint8, int8, etc., are not defined in the B Model. But we only need to limit the range of values on B types FLOAT or NAT when translating, usually, we define the data type name under *CONSTANTS* and restrict the range of data types under *PROPERTIES*. For example.

```
CONSTANTS
    uint32_t,
    uint16_t,
    uint8_t,
    int32_t,
    int16_t,
    int8_t

PROPERTIES
    uint32_t = 0..0xFFFFFFFF ∧
    uint16_t = 0..0xFFFF ∧
    uint8_t = 0..0xFF ∧
    int32_t = -2147483648..2147483647 ∧
    int16_t = -32768..32767 ∧
    int8_t = -128..127
```

Figure 2. Data type definition for B model

The syntax paradigm and examples for defining enumeration types in SCADE are shown in Figure 3.

```
type_def ::= type_expr
           | enum { ID {{ , ID }} }
```

```
type
MOVE = enum {Stop, Forward, Reverse};
DIRECTION = enum {North, South, East, West};
```

Figure 3. Enumeration type definition of SCADE

There is no enumeration of types in the B model. Enumerated types can be translated as *SETS* of B models. Figure 4 shows the syntax paradigm and examples of the definition *SETS* of the B model.

```
SETS
  type_def ::= { ID {{ , ID }} }
```

```
SETS
  MOVE = {Stop, Forward, Reverse} ;
  DIRECTION = {North, South, East, West}
```

Figure 4. Set definition of the B model

The syntax paradigm and examples for defining array types in SCADE are shown in Figure 5.

```
array_expr ::= expr ˆ expr
             | [ list ]
```

```
a1: uint32^3;
a2: bool^3;
```

Figure 5. SCADE array type definition

In SCADE, we use square brackets to take any value from the array, such as *a1[1]*, or assign, such as *a1[1] = 10*. In the B model, the symbol "→" is used to represent the full function relationship, and the domain of this relationship is required to be continuous, as shown in Figure 6.

```
array_expr ∈ expr → expr
```

$$a1 \in 0..2 \to uint8\_t \land$$
$$a2 \in 0..2 \to BOOL$$

Figure 6. Array type definition for the B model

In the B model, we can take any value from the array by using parentheses, such as *a1(1)*, or assign, such as *a1(1) := 10*.

The structure of SCADE defines the syntax format and examples as shown in Figure 7. Elements of a structure can be accessed by, for example, *tstr.l1 = 10, tstr.l2 [0] = 10*.

```
struct_expr ::= expr . ID
              | {label_expr {{ , label_expr }} }
label_expr ::= ID : expr
```

```
type
Tstr = {l1: int32, l2: int32^2};
```

Figure 7. Structure type definition for SCADE

The specific syntax paradigm of structure in the B model is shown in Figure 8.

```
Set_of_records ::= struct(ID ∈ expr (( , ID ∈ expr )) )
```

$$Tstr = struct(l1 \in int32\_t, l2 \in 0..1 \to int32\_t)$$

Figure 8. Structure type definition for B model

The elements of a structure can be accessed, for example, using *Tstr'l1 := 10, Tstr'l2(0) := 10*. Alternatively, you can assign values to the entire structure using rec. For example: *Tstr = rec(10, (20, 30))*.

3.2 Constant

A constant is always available and has the same value throughout the program execution. The definition syntax paradigm and examples of SCADE constants are shown in Figure 9.

```
const_block ::= const {{ const_decl ; }}
  const_decl ::= interface_status ID : type_expr [[ = expr ]]
```

```
const p1: uint8 = 100;
const p2: bool = true;
```

Figure 9. Constant type definition for SCADE

In the B model, you need to specify types via *CONSTANTS* and set values via *PROPERTIES*. The syntax paradigm and examples are shown in Figure 10.

```
CONSTANTS
  ID ∈ type_expr

PROPERTIES
  ID := expr
```

```
CONSTANTS
  p1 ∈ uint8_t ∧
  p2 ∈ BOOL

PROPERTIES
  p1 := 100 ∧
  p2 := TRUE
```

Figure 10. Constant type definition for B model

3.3 Equation

Equations allow to definition of the data flow expression associated with an output or local identifier. This expression is evaluated at each cycle, according to input/output and current/previous values, and then assigned to this identifier. According to the problem at hand, equations can either be declared in a data flow or a control-flow flavor. Its syntax paradigm is shown in Figure 11.

```
equation ::= lhs = expr
      lhs ::= ( )
              | lhs_id {{ , lhs_id }}
   lhs_id ::= ID
```

Figure 11. SCADE's syntax paradigm for equations

Assign an expression to a list of identifiers, putting in front the data flow aspect of this expression, such as Basic Expressions, Boolean Expressions Arithmetic Expressions, Relational Expressions, and so on. The equations of the B model are similar to those of the SCADE model, as shown in Figure 12, except that ":=" is used instead of "=".

```
L3 = L1 + L2;
L5 = L4 * L3;
```

```
L3 := L1 + L2;
L5 := L4 * L3
```

Figure 12. SCADE's equations and B's equations

3.4 Conditional Expression

Conditional Expressions are convenient to express control structures when the control flow only depends on a condition computable in the current cycle. Depending on its type (Boolean or enumerated), this condition may lead to two or more switch cases. Each case proposes a definition of a subset of the whole set of variables defined by this conditional block. Undefined variables are either maintained to their previous value (the last one) or follow a default behavior stated in their declaration, Conditional expressions of SCADE are divided into IF expressions and CASE expressions. The syntax paradigm and examples are shown in Figure 13.

```
switch_expr ::= if expr then expr else expr
              | ( case expr of {{ case_expr }}+ )
case_expr ::= | pattern : expr
pattern ::= path_id
          | CHAR
          | [[ – ]] INTEGER
          | [[ – ]] TYPED_INTEGER
          | bool_atom
          | _
```

```
L8 = if L7 then (L5) else (L6);
```

```
L1 = ( case L5 of
     | 1 :  L2
     | 2 :  L3
     | _ :  L4);
```

Figure 13. Conditional expressions for SCADE

The IF conditional expression syntax paradigm and example of the B model are shown in Figure 14.

```
If_substitution ::=
"IF" Predicate "THEN" Substitution
[ "ELSIF" Predicate "THEN" Substitution ]*
[ "ELSE" Substitution ]
"END"
```

```
IF L7 = TRUE THEN
   L8 := L5
ELSE
   L8 := L6
END ;
```

Figure 14. IF condition expression of B model

Figure 15 shows the syntax paradigm and example of CASE condition expression of the B model.

```
Case_substitution ::=
"CASE" Expression "OF"
"EITHER" Simple_term+ "," "THEN" Substitution
( "OR" Simple_term+ "," "THEN" Substitution )+
[ "ELSE" Substitution ]
"END"
"END"
```

```
CASE L5 OF
EITHER 1 THEN
   L1 := L2
OR 2 THEN
   L1 := L3
ELSE
   L1 := L4
END
END
```

Figure 15. CASE condition expression for B model

3.5 Sequential Operator

Operator *pre* is a sequential primitive that Shifts Flows on the last instant backward when this flow was defined within the same scope. It thus produces an undefined value at its first instant of activation called nil. The "→" primitive evaluates its left argument at its first instant of evaluation or after a restart, and its right argument otherwise.

The flow *fby(b;n;a)* combines the first two primitives in order to access previous values and produce only well-initialized flows. It can be equivalently defined by:

$$fby(b; n; a) = a \rightarrow pre\ fby(b; n-1; a) = \underbrace{a \rightarrow pre\ (a \rightarrow pre\ (...(a \rightarrow pre\ b)...))}_{n\ times}$$

It then evaluates its third argument on the first n cycles (or the first n cycles after a restart), and its first one on the remaining cycles.

There is no *fby* in the B model, so we need to build the B model with the same functionality as *fby*, the most critical of which is to build the data structure for storing previous values and the operation for Shifts Flows. In *VARIABLES*, we can define the variable used to store all data in Shifts Flows. In *INVARIANT* define the type of the variable, which should be an array to store the data for cycles over Shifts Flows, initialize the array in *INITIALISATION*, and define the behavior of Shifts Flows in *OPERATIONS*. Figure 16 shows an and operator *c=fby(b;n;a)* Equivalent B model.

```
VARIABLES
    store

INVARIANT
    store ∈ 0..(n-1) → type_expr

INITIALISATION
    store := (0..(n-1))*{a};

OPERATIONS
    c ← fby(b) =
        BEGIN
            c := store(0);
            store(0) := store(1);
            ……
            store(n-2) := store(n-1);
            store(n-1) := b
        END
    END
```

Figure 16. Sequential logic of B model

3.6 State Machine

State Machines offer a model's most elaborate combination of control and data flow information. Intuitively, State Machines extend the conditional block construct when the condition cannot be computed without memorizing extra expressions. In this case, the control structure can be best expressed by means of a State Machine whose states contain the needed information. The syntax paradigm of SCADE for state machine is shown in Figure 17.

```
state_machine ::= automaton [[ ID ]] {{ state_decl }}
    state_decl ::= [[ initial ]] [[ final ]] state ID
                   [[ unless {{ transition ; }}+ ]]
                   data_def
                   [[ until {{ transition ; }} ]]
                   [[ synchro [[ actions ]] fork ; ]] ]]

    transition ::= if expr arrow
        arrow ::= [[ actions ]] fork
         fork ::= target
                | if expr arrow {{ elsif_fork }} [[ else_fork ]]
endelsif_fork ::= elsif expr arrow
    else_fork ::= else arrow
       target ::= restart ID
                | resume ID

      actions ::= do { [[ emit ]] emission_body
                       {{ ; [[ emit ]] emission_body }} }
                | do data_def
```

Figure 17. State Machine Syntax Paradigm for SCADE

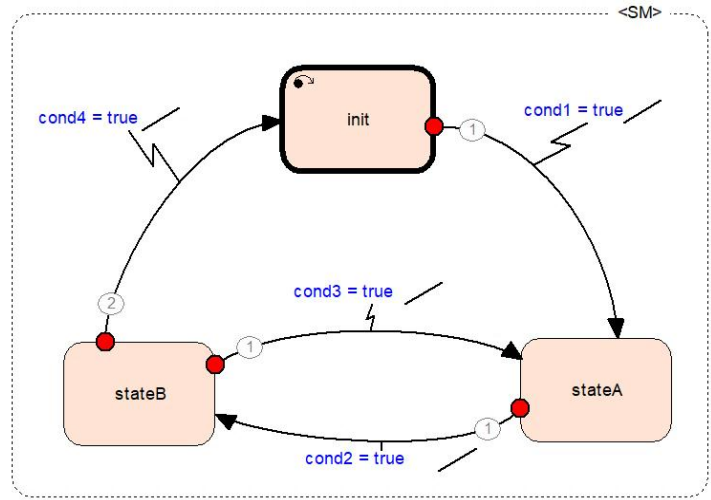

Figure 18. Example of a SCADE state machine

As shown in Figure 18, the SCADE state machine in the SCADE Suite graphical interface has three states *init, stateA, stateB*, and three transition conditions *cond1, cond2, and cond3.* The initial state is *init*. When *cond1=true*, the state is transferred from *init* to *stateA*. When *cond2=true*, the state will be migrated from *stateA* to *stateB*; when *cond3=true*, the state will be migrated from *stateB* to *stateA*. Textual syntax paradigm as shown in Figure 17 is used to describe the state machine, as shown in Figure 19

```
automaton SM
  initial state init
  unless
    if cond1 = true
      restart stateA;

  state stateB
  unless
    if cond3 = true
      restart stateA;
    if cond4 = true
      restart init;

  state stateA
  unless
    if cond2 = true
      restart stateB;
returns .. ;
```

Figure 19. Example of a textual SCADE state machine

In model B, in order to describe such a state machine, we can define all states in the state machine in *SETS*, and we specify all states with enumerated sets. In *INVARIANT*, define a variable in *VARIABLES* to represent the current state of the state machine. In *INVARIANT*, specify the type of the variable, which should be an element of the set. Initialize this variable in *INITIALISATION* to the initial state of the state machine. We use the *CASE* condition expression as shown in Figure 15 to describe each state of the state machine. In any state, as long as the transition condition is satisfied, it will jump to the next state; IF not, it will keep the original state. Therefore, we use the IF condition expression as shown in Figure 14 to represent the migration relationship of this state in each state.

The state machine shown in Figure 18 and Figure 19 is described by model B as shown in Figure

20.

```
SETS
  STATE = {init, stateA, stateB}

VARIABLES
  sm_state

INVARIANT
  sm_state ∈ STATE

INITIALISATION
  sm_state := init

OPERATIONS
  SM(cond1, cond2, cond3, cond4) =
    PRE
      cond1 ∈ BOOL ∧ cond2 ∈ BOOL ∧ cond3 ∈ BOOL ∧ cond4 ∈ BOOL
    THEN
      CASE sm_state OF
        EITHER init THEN
            IF cond1 = TRUE THEN
              sm_state := stateA
            END
        OR stateA THEN
            IF cond2 = TRUE THEN
              sm_state := stateB
            END
        OR stateB THEN
            IF cond3 = TRUE THEN
              sm_state := stateA
            ELSIF cond4 = TRUE THEN
              sm_state := init
            END
        END
      END
    END
  END
```

Figure 20. An example of building a state machine with the B model

3.7 Higher-Order Operator

A user-defined operator can be used in its scope as any primitive operator to build expressions. Scade also provides higher-order primitives that modify the behavior of operators: They take as input an operator and return another operator that can thus be applied as usual operators.

Higher-Order includes *map, mapi, mapw, mapwi, fold, foldi, foldw, foldwi, mapfold, mapfoldi, mapfoldw, mapfoldwi*.

There is no such high-order operator in the B model. High-order operators are mainly used to process data such as arrays and need to iterate over the same operator. Therefore, we use the *WHILE* loop in the B model to construct such high-order operators. Meanwhile, some local variables need to be used in the *WHILE* loop. The syntax paradigm of the *WHILE* loop and local variables of the B model are shown in Figure 21.

```
While_substitution ::=
"WHILE" Condition "DO" Instruction
"INVARIANT" Predicate
"VARIANT" Expression
"END"
```

```
Var_substitution ::= "VAR" Ident+"," "IN" Substitution "END"
```

Figure 21. WHILE loop syntax paradigm and local variables syntax of B model

3.7.1 map

Let *op* be an operator taking *n* parameters as input and producing *k* output values. Let $A_1,...,A_n$ be arrays of size *size* having the corresponding basic types as the inputs of operator *op*. Then $v_1, ..., v_m$ such that:

$$v_1, ..., v_m = (map\ op\ <<size>>)(A_1, ..., A_n)$$

are arrays of size defined by:

$$\forall i \in [0...size], v_1[i],...,v_m[i]=op(A_1[i],...,A_n[i])$$

The structure of *map* constructed through the *WHILE* loop is shown in Figure 21, where *VAR* defines a local variable *idx*, which is used to iterate over every element in the number group. The range of *idx* should be $0 \leq idx < size$. The body *DO* clause performs *op* on each element of the array, so the *DO* clause should be:

$$v_1(idx), ..., v_m(idx) \leftarrow op(A_1(idx), ..., A_n(idx)).$$

*idx* should also be added after *op* execution.

Since the *WHILE* loop may execute multiple loop bodies and experience a series of intermediate states, which must be different from each other, in order to ensure the correctness of the *WHILE* loop, a loop invariant is needed to describe the relationship from the initial state to the termination state, and to describe the relationship, these intermediate states must be involved. The B model uses *INVARIANT* to express loop invariant and a fixed predicate to describe the changing state of loop execution.

The higher-order Operator executes *op* every time, which brings great convenience for us to design loop invariant. If we use S to represent the functional specification of *op*, all completed loops should conform to the functional specification of *op*. Therefore, the loop invariant can be designed to be:

$$\forall i.(i \in (0 .. idx) \Rightarrow S)$$

The loop invariant can only describe a certain final state that the execution of the loop can reach, but it cannot guarantee that the loop can reach the final state, and the problem of whether the loop can be terminated must be considered separately. The B model introduces *VARIANT*, which is an expression with values of natural numbers. To prove that a loop must terminate, we need to find a natural numerical expression that ensures that the execution of the loop is monotonically decreasing. Here obviously *size - idx* can be expressed.

The final B model equivalent to *map* is shown in Figure 22.

```
VAR idx IN
    idx := 0;
    WHILE idx < size
    DO
        v₁(idx), ..., vₘ(idx) ← op(A₁(idx), ..., Aₙ(idx));
        idx := idx + 1
    INVARIANT
        ∀i.(i ∈ (0 .. idx) ⇒ S)
    VARIANT
        size - idx
    END
END
```

Figure 22. Equivalent to the B model of *map* operator

If you want to add two arrays, for example：

$A_1 = [1,2,3,4,5,6,7,8,9,10]$

$A_2 = [0,1,2,3,4,5,6,7,8,9]$

The result will be

v = [1,3,5,7,9,11,13,15,17,19]

To do this calculation with SCADE, use this expression: $v = (map\ op\ <<10>>)(A_1, A_2)$; The calculation is completed by the B model as shown in Figure 23.

```
VAR idx IN
    idx := 0;
    WHILE idx < 10
    DO
        v(idx) ← op(A₁(idx), A₂(idx));
        idx := idx + 1
    INVARIANT
        ∀i.(i ∈ (0 .. idx) ⇒ v(i) = A₁(i) + A₂(i))
    VARIANT
        10 - idx
    END
END
```

Figure 23. Example of B model equivalent to *map* operator

3.7.2 mapi

*mapi* behaves as *map*, but operator *op* is required to take an extra integer argument as its first input. The current iteration index is passed as this first argument:

$$\forall i \in [0...size],\ v_1[i],...,v_m[i]=op(i,A_1[i],...,A_n[i])$$

Figure 24 shows the B model equivalent to *mapi*.

```
VAR idx IN
    idx := 0;
    WHILE idx < size
    DO
        v₁(idx), ..., vₘ(idx) ← op(idx, A₁(idx),..., Aₙ(idx));
        idx := idx + 1
    INVARIANT
        ∀i.(i ∈ (0 .. idx) ⇒ S)
    VARIANT
        size - idx
    END
END
```

Figure 24. Equivalent to the B model of *mapi* operator

3.7.3 fold

Let *op* be an operator taking *n+1* parameters as input and producing one output value of the same type as its first input. Let $A_1,...,A_n$ be arrays of *size* having the corresponding basic types as the inputs of operator *op*, and *exp* be an expression of the first input type. Then *acc* such that:

$$acc = (fold\ op\ <<size>>)\ (exp, A_1, ..., A_n)$$

is an expression of the same type as exp defined by:

$$\begin{cases} acc = acc_{size} \\ \forall i \in [0...size[, acc_{i+1} = op(acc_i, A_1[i],...,A_n[i]) \\ acc_0 = expr \end{cases}$$

Figure 25 shows the B model equivalent to *fold*.

```
VAR idx, acc IN
  idx, acc := 0, expr;
  WHILE idx < size
    DO
      acc, v_1(idx), ..., v_m(idx) ← op(acc, A_1(idx), ..., A_n(idx));
      idx := idx + 1
    INVARIANT
      ∀i.(i ∈ (0 .. idx) ⇒ S)
    VARIANT
      size - idx
  END
END
```

Figure 25. Equivalent to the B model of *fold* operator

3.7.4 foldi

*foldi* behaves the same as fold, but operator *op* is required to take an extra integer argument as its first input. The current iteration index is passed as this first argument:

$$\begin{cases} acc = acc_{size} \\ \forall i \in [0...size[, acc_{i+1} = op(i, acc_i, A_1[i], ..., A_n[i]) \\ acc_0 = expr \end{cases}$$

Figure 26 shows the B model equivalent to *foldi*.

```
VAR idx, acc IN
  idx, acc := 0, expr;
  WHILE idx < size
    DO
      acc, v_1(idx), ..., v_m(idx) ← op(idx, acc, A_1(idx), ..., A_n(idx));
      idx := idx + 1
    INVARIANT
      ∀i.(i ∈ (0 .. idx) ⇒ S)
    VARIANT
      size - idx
  END
END
```

Figure 26. Equivalent to the B model of *foldi* operator

3.7.5 mapfold

Let *op* be an operator taking $a+n$ parameters as input and producing $a+m$ output values, such that their a first item have the same type. Let $A_1,...,A_n$ be arrays of size *size* having the corresponding basic types as the inputs of operator *op*, and $exp^1, ..., exp^a$ be expressions of these first a items types. The equation:

$$acc^1, ..., acc^a, v_1, ..., v_m = (mapfold\ a\ op\ <<size>>)(exp^1, ..., exp^a, A_1, ..., A_n)$$

is equivalent to:

$$\begin{cases} \forall j \in [1...a], acc^j = acc^j_{size} \\ \forall i \in [0...size[, acc^1_{i+1}, ..., acc^a_{i+1}, v_1[i], ..., v_m[i] = op(acc^1_i, ..., acc^a_i, A_1[i], ..., A_n[i]) \\ \forall j \in [1...a], acc^j_0 = exp^j \end{cases}$$

Figure 27 shows the B model equivalent to *mapfold*.

```
VAR idx, acc¹,..., accᵃ, IN
  idx, acc¹,..., accᵃ := 0, expr¹,..., exprᵃ;
WHILE idx < size
  DO
    acc¹,..., accᵃ, v₁(idx), ..., vₘ(idx) ← op(acc¹,..., accᵃ, A₁(idx), ..., Aₙ(idx));
    idx := idx + 1
  INVARIANT
    ∀i.(i ∈ (0 .. idx) ⇒ S)
  VARIANT
    size - idx
  END
END
```

Figure 27. Equivalent to the B model of *mapfold* operator

3.7.6 mapfoldi

Let *op* be an operator taking $a+n+1$ parameters as input and producing $a+m$ output values. Let $A_1, ..., A_n$ be arrays of size *size* having the corresponding basic types as the inputs of operator *op*, and $exp^1, ..., exp^a$ be an expression of these first items type. The equation:

$acc^1, ..., acc^a, v_1, ..., v_m = $ *(mapfoldi a op <<size>>)*$(exp^1, ..., exp^a, A_1, ..., A_n)$

is equivalent to:

$$\begin{cases} \forall j \in [1...a], acc^j = acc^j_{size} \\ \forall i \in [0...size[, acc^1_{i+1},...,acc^a_{i+1}, v_1[i],...,v_m[i] = op(i, acc^1_i,...,acc^a_i, A_1[i],...A_n[i]) \\ \forall j \in [1...a], acc^j_0 = exp^j \end{cases}$$

Figure 28 shows the B model equivalent to *mapfoldi*.

```
VAR idx, acc¹,..., accᵃ IN
  idx, acc¹,..., accᵃ := 0, expr¹,..., exprᵃ;
WHILE idx < size
  DO
    acc¹,..., accᵃ, v₁(idx), ..., vₘ(idx) ← op(idx, acc¹,..., accᵃ, A₁(idx), ..., Aₙ(idx));
    idx := idx + 1
  INVARIANT
    ∀i.(i ∈ (0 .. idx) ⇒ S)
  VARIANT
    size - idx
  END
END
```

Figure 28. Equivalent to the B model of *mapfoldi* operator

3.7.7 mapw

Let *op* be an operator taking *n* parameters as input and producing $k+1$ output values, its first output being a Boolean expression. Let $A_1, ..., A_n$ be arrays of size *size* having the corresponding basic types as the inputs of operator *op*, *initcond* a Boolean expression, and $d_1, ..., d_m$ some default values of the same type as the outputs of *op*. The equation:

$idx, v_1, ..., v_m = $ *(mapw op <<size >>*
    *if   initcond*
    *default   (d₁, ..., dₘ)(A₁, ..., Aₙ)*;

*is equivalent to:*

$$\begin{cases} cond_0 = initcond \\ \forall i \in [0\ldots idx[,(cond_{i+1}, v_1[i],\ldots,v_m[i]) = op(A_1[i],\ldots,A_n[i]) \\ \forall i \in [0\ldots idx-1[, cond_i = \mathbf{true} \\ idx = size \lor cond_{idx-1} = \mathbf{false} \\ \forall i \in [idx\ldots size[, cond_i = cond_{i-1} \\ \forall i \in [idx\ldots size[, \forall j \in [1, m], v_j[i] = d_j \end{cases}$$

Here we must note that *cond* is the condition for starting and ending iterations, and when translated to B model, *cond = TRUE* must be put into *WHILE* condition and *INVARIANT*. Figure 29 shows the B model equivalent to *mapw*.

```
VAR idx, cond IN
    idx, cond := 0, initcond;
    WHILE idx < size ∧ cond = TRUE
    DO
        cond, v₁(idx), ..., vₘ(idx) ← op(A₁(idx), ..., Aₙ(idx));
        idx := idx + 1
    INVARIANT
        ∀i.((i ∈ (0 .. idx) ∧ cond = TRUE) ⇒ S)
    VARIANT
        size - idx
    END
END
```

Figure 29. Equivalent to B model of *mapw* operator

### 3.7.8 mapwi

*mapwi* behaves the same as *mapw*, but operator *op* is required to take an extra integer argument as its first input. The current iteration index is passed as this first argument:

$$idx, v_1, \ldots, v_m = (mapwi\ op\ <<size>>$$
$$if\ \ initcond$$
$$default\ \ (d_1, \ldots, d_m)(A_1, \ldots, A_n);$$

is equivalent to:

$$\begin{cases} cond_0 = initcond \\ \forall i \in [0\ldots idx[,(cond_{i+1}, v_1[i],\ldots,v_m[i]) = op(i, A_1[i],\ldots,A_n[i]) \\ \forall i \in [0\ldots idx-1[, cond_i = \mathbf{true} \\ idx = size \lor cond_{idx-1} = \mathbf{false} \\ \forall i \in [idx\ldots size[, cond_i = cond_{i-1} \\ \forall i \in [idx\ldots size[, \forall j \in [1, m], v_j[i] = d_j \end{cases}$$

Figure 30. shows the B model equivalent to *mapwi*.

```
VAR idx, cond IN
  idx, cond := 0, initcond;
  WHILE idx < size ∧ cond = TRUE
  DO
    cond, v_1(idx), ..., v_m(idx) ← op(idx, A_1(idx), ..., A_n(idx));
    idx := idx + 1
  INVARIANT
    ∀i.((i ∈ (0 .. idx) ∧ cond = TRUE) ⇒ S)
  VARIANT
    size - idx
  END
END
```

Figure 30. Equivalent to the B model of *mapwi* Operator

3.7.9 foldw

Let *op* be an operator taking $n+1$ parameters as input and producing two output values, a Boolean and an output having the same type as the first input. Let $A_1, ..., A_n$ be arrays of size *size* having the corresponding basic types as the inputs of operator *op*, and $acc_0$ be an expression of the first input type. Then *idx* and *acc* such that:

$$idx, acc = (foldw\ op\ <<size>>\ if\ cond_0)(acc_0, A_1, ..., A_n)$$

are defined by $idx=idx_{size}$ and $acc=acc_{size}$ such that:

$$\forall i \in [0...size[, idx_{i+1}, cond_{i+1}, acc_{i+1} = \begin{cases} idx_i + 1, op(acc_i, A_1[i],...,A_n[i])\ if\ cond_i = \textbf{true} \\ idx_i, acc_i, cond_i\ otherwise \end{cases}$$

Figure 31. shows the B model equivalent to *foldw*.

```
VAR idx, acc, cond IN
  idx, acc, cond := 0, expr, initcond;
  WHILE idx < size ∧ cond = TRUE
  DO
    acc, cond, v_1(idx), ..., v_m(idx) ← op(acc, A_1(idx), ..., A_n(idx));
    idx := idx + 1
  INVARIANT
    ∀i.((i ∈ (0 .. idx) ∧ cond = TRUE) ⇒ S)
  VARIANT
    size - idx
  END
END
```

Figure 31. Equivalent to B model of *foldw* Operator

3.7.10 foldwi

*foldwi* behaves the same as *foldw*, but operator *op* is required to take an extra integer argument as its first input. The current iteration index is passed as this first argument:

$$\forall i \in [0...size[, idx_{i+1}, cond_{i+1}, acc_{i+1} = \begin{cases} idx_i + 1, op(\ acc_i, A_1[i],...,A_n[i])\ if\ cond_i = \textbf{true} \\ idx_i, acc_i, cond_i\ otherwise \end{cases}$$

Figure 32 shows the B model equivalent to *foldwi*

```
VAR idx, acc, cond IN
    idx, acc, cond := 0, expr, initcond;
WHILE idx < size ∧ cond = TRUE
    DO
        acc, cond, v_1(idx), ..., v_m(idx) ← op(idx, acc, A_1(idx), ..., A_n(idx));
        idx := idx + 1
    INVARIANT
        ∀i.((i ∈ (0 .. idx) ∧ cond = TRUE) ⇒ S)
    VARIANT
        size - idx
    END
END
```

Figure 32. Equivalent to the B model of *foldwi* Operator

3.7.11 mapfoldw

Let *op* be an operator taking $a+n$ parameters as input and producing $a+m+1$ output values, such that their first items have the same type. Let $A_1, ..., A_n$ be arrays of size *size* having the corresponding basic types as the inputs of operator *op*, and $exp^1, ..., exp^a$ be expression of these first a items type. The equation:

$idx, cond, acc^1, ..., acc^a, v_1, ..., v_m =$

(*mapfoldw a op << size >> if initcond default* $(d_1, ..., d_m))(exp^1, ..., exp^a, A_1, ..., A_n)$

is equivalent to:

$$\begin{cases} cond_0 = initcond \\ \forall j \in [1...a], acc_0^j = exp^j \\ \forall i \in [0...idx[, cond_{i+1}, acc_{i+1}^1, ..., acc_{i+1}^a, v_1[i], ..., v_m[i] = op(acc_i^1, ..., acc_i^a, A_1[i], ..., A_n[i]) \\ \forall i \in [0...idx-1[, cond_i = \textbf{true} \\ idx = size \vee cond_{idx-1} = \textbf{false} \\ \forall i \in [idx...size[, cond_i = cond_{i-1} \\ \forall i \in [idx...size[, \forall j \in [1, m] v_j[i] = d_j \\ \forall i \in [idx...size[, \forall j \in [1...a] acc_i^j = acc_{i-1}^j \\ cond = cond_{size} \\ \forall j \in [1...a], acc^j = acc_{size}^j \end{cases}$$

Figure 33 shows the B model equivalent to *mapfoldw*.

```
VAR idx, acc^1, ..., acc^a, cond IN
    idx, acc^1, ..., acc^a, cond := 0, expr^1, ..., expr^a, initcond;
WHILE idx < size ∧ cond = TRUE
    DO
        acc^1, ..., acc^a, cond, v_1(idx), ..., v_m(idx) ← op(acc^1, ..., acc^a, A_1(idx), ..., A_n(idx));
        idx := idx + 1
    INVARIANT
        ∀i.((i ∈ (0 .. idx) ∧ cond = TRUE) ⇒ S)
    VARIANT
        size - idx
    END
END
```

Figure 33. Equivalent to the B model of *mapfoldw* Operator

3.7.12 mapfoldwi

Let *op* be an operator taking *a+n+1* parameters as input and producing *a+m+1* output values, such that their first items have the same type. Let $A_1, ..., A_n$ be arrays of size *size* having the corresponding basic types as the inputs of operator *op*, and $exp^1, ..., exp^a$ be the expression of these first an items type. The equation:

$idx, cond, acc^1, ..., acc^a, v_1, ..., v_m =$

$(mapfoldwi\ a\ op << size >> if\ initcond\ default\ (d_1, ..., d_m))(exp^1, ..., exp^a, A_1, ..., A_n)$

is equivalent to:

$$\begin{cases} cond_0 = initcond \\ \forall j \in [1...a], acc_0^j = exp^j \\ \forall i \in [0...idx[, cond_{i+1}, acc_{i+1}^1, ..., acc_{i+1}^a, v_1[i], ..., v_m[i] = op(i, acc_i^1, ..., acc_i^a, A_1[i], ..., A_n[i]) \\ \forall i \in [0...idx-1[, cond_i = \textbf{true} \\ idx = size \vee cond_{idx-1} = \textbf{false} \\ \forall i \in [idx...size[, cond_i = cond_{i-1} \\ \forall i \in [idx...size[, \forall j \in [1, m], v_j[i] = d_j \\ \forall i \in [idx...size[, \forall j \in [1...a], acc_i^j = acc_{i-1}^j \\ cond = cond_{size} \\ \forall j \in [1...a], acc^j = acc_{size}^j \end{cases}$$

Figure 34. shows the B model equivalent to *mapfoldwi*.

```
VAR idx, acc¹,..., accᵃ, cond IN
  idx, acc₁,..., accₐ, cond := 0, expr¹,..., exprᵃ, initcond;
WHILE idx < size ∧ cond = TRUE
  DO
    acc¹,..., accᵃ, cond, v₁(idx), ..., vₘ(idx) ← op(idx, acc¹,..., accᵃ, A₁(idx), ..., Aₙ(idx));
    idx := idx + 1
  INVARIANT
    ∀i.((i ∈ (0 .. idx) ∧ cond = TRUE) ⇒ S)
  VARIANT
    size - idx
  END
END
```

Figure 34. Equivalent to B model of *mapfoldwi* Operator

## 4. Experiment 1

This experiment demonstrates in detail how various components of a SCADE model, including data types, constants, operators, state machines, and conditional expressions, are translated into equivalent B-Model constructs. By constructing a sample SCADE model that encompasses multiple modeling elements, this experiment verifies the applicability of the proposed translation method.

To validate the correctness of the translation, the experiment compares the simulation results of the original SCADE model and its corresponding B-Model. By using the same input values for simulation, the comparison ensures that the states and outputs of both models remain identical in every execution cycle, thereby proving the correctness of the SCADE-to-B translation process.

Furthermore, by converting the SCADE model into a B-Model and utilizing the ProB Animator and Model Checker for simulation verification, this experiment introduces a new approach to verifying

SCADE models using ProB. This method enhances the verification capabilities of SCADE models, providing additional support for model analysis.

The following sections will demonstrate the translation process from a SCADE model to a B-Model and compare the simulation execution results of both models to verify the correctness of the translation. The SCADE model constructed for this experiment is not designed to implement a specific real-world function but rather to incorporate as many key modeling components as possible, including Type, Constant, Equation, Condition Expression, Sequential Operator, State Machine, Higher-Order Operator.

The constructed SCADE model is shown below. It includes Constants, Types, and Operators, where the Operators section contains a ComputeSum model. The inputs and outputs of this model are defined in the Interface, including input, fby_in (previous value input), output, fby_out (previous value output), and strucDemo (structured data example). Additionally, the model features a state machine STATE and a conditional block IfBlock, which contribute to the verification of the translation approach.

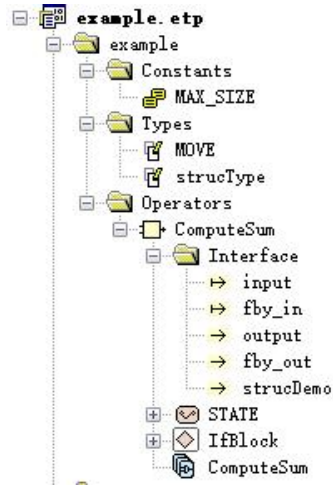

Figure 35. The structure of the SCADE Suite project

The type definitions in the model are as follows. They include an enumeration type MOVE and a structured type strucType.

Figure 36. Type definition of SCADE project

The constant MAX_SIZE is defined as shown below.

Figure 37. Constant definition of the SCADE Suite project

The model also includes the following components:
- A Higher-Order Operator "map": This operator performs a multiplication operation on the input array input and then outputs the result to output.

- A Sequential Operator "FBY": This operator delays the input *fby_in* by 3 cycles before passing it to the output *fby_out*.
- A State Machine STATE: The state machine consists of three states, *init*, *stateA*, and *stateB*, along with three transition conditions.
- A Conditional Expression (IF expression): Within this conditional expression, a structure *strucDemo* of type *strucType* is assigned a value.

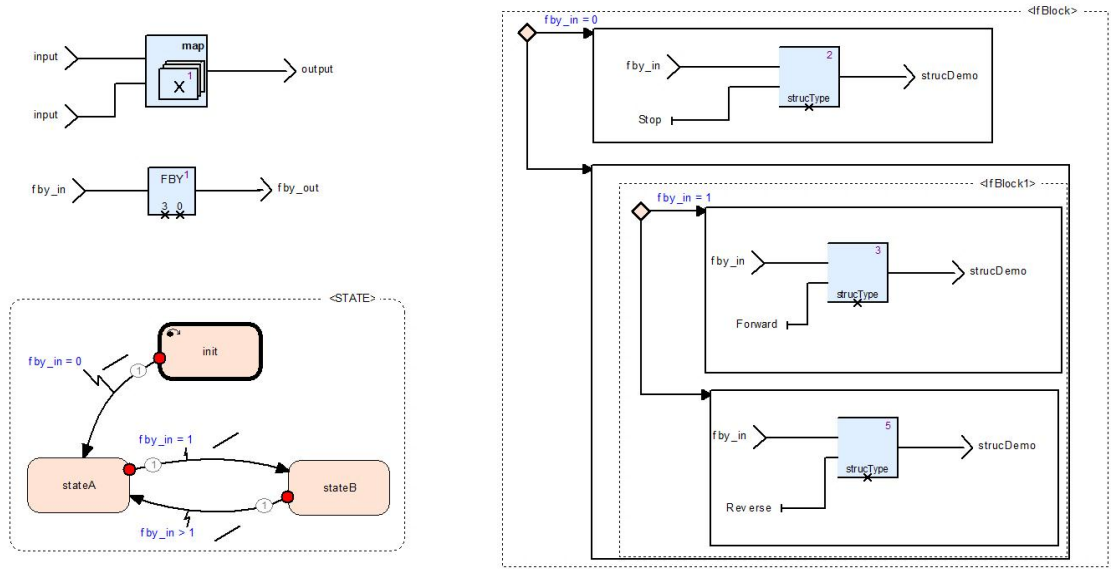

Figure 38. State Machines and Conditional Expressions in SCADE Suite

The figure above shows the graphical interface of SCADE, and the corresponding text-based SCADE language model is provided in Appendix 1. In the text-based SCADE model shown in Appendix 2, all the SCADE model elements from the figure can be observed.

As shown in Appendix 2, the B-Model has been generated by applying the translation rules to the SCADE model. Next, we will compare the models before and after translation and validate the correctness of the translation through simulation.

For simulation, the SCADE model is simulated using the ANSYS SCADE Suite, while the B-Model is simulated using ProB. The comparison is performed by applying the same input to both models and verifying whether their internal states and outputs remain consistent.

The following is an explanatory translation of the above models.

A. The SCADE data type *uint8* and constant *MAX_SIZE* are translated into the B-Model as follows:

```
CONSTANTS
    uint8_t, MAX_SIZE

PROPERTIES
    uint8_t = 0..255 ∧
    MAX_SIZE = 5
```

B. The enumeration type *MOVE* in SCADE is translated into a set in the B-Model.

```
SETS
    MOVE = {Stop, Forward, Reverse};
```

C. The array input input in SCADE is represented as in the B-Model:

$$\text{input} \in 0..(MAX\_SIZE - 1) \rightarrow uint8\_t$$

D. The assignment of the structure *strucDemo* in SCADE is represented in the B-Model using rec for assignment.

$$strucDemo := rec(fby\_data \in fby\_in, move \in Forward)$$

E. Many equations "=" in SCADE are translated into assignment statements ":=" in the B-Model.

F. The if expression in SCADE is translated into an *IF* expression in the B-Model.

G. In the B-Model translated from the *fby* sequential operator in the SCADE model:

```
VARIABLES
    store
INVARIANT
    store ∈ 0..2 → uint8
INITIALISATION
    store := { 0 ↦ 0, 1 ↦ 0, 2 ↦ 0 }
```

An array *store* is defined to store the data of *fby* for each cycle and is initialized to 0 in the operation *ComputeSum*:

```
fby_out := store(0);
store(0) := store(1);
store(1) := store(2);
store(2) := fby_in;
```

Similar to a FIFO, the values are shifted sequentially to implement the behavior of *fby*.

H. The state machine in SCADE is translated into the following B-Model, where:

```
SETS
    STATE = {init, stateA, stateB}

VARIABLES
    sm_state

INVARIANT
    sm_state ∈ STATE

INITIALISATION
    sm_state := init
```

The sm_*state* is used to define the state space and set the initial state to init. The *CASE* statement, as shown in the figure below, represents the state transition relationships of the state machine.

```
CASE sm_state OF
  EITHER init THEN
    IF fby_in = 0 THEN
      sm_state := stateA
    END
  OR stateA THEN
    IF fby_in = 1 THEN
      sm_state := stateB
    END
  OR stateB THEN
    IF fby_in > 1 THEN
      sm_state := stateA
    END
  END
END
```

I. The *map* operator in the SCADE model is translated into the following statement in the B-Model.

```
VAR idx IN
  idx := 0;
  WHILE idx < MAX_SIZE DO
    output(idx) := input(idx) * input(idx);
    idx := idx + 1
  INVARIANT
    ∀i.(i ∈ 0..(idx - 1) ⇒ output(i) = input(i) * input(i))
  VARIANT
    MAX_SIZE - idx
  END
END
```

The following is the simulation results comparison.

A. By running a simulation in ANSYS SCADE Suite for the SCADE model and using ProB to simulate the B-Model, we compare the states and outputs of both models under the same input conditions.

In the first cycle, we set all inputs to 0 and perform a step-by-step execution. The results are shown in the figures below. Figure 39 is the simulation result of the SCADE model. Figure 40 and Figure 41 are the simulation results of the B-Model.

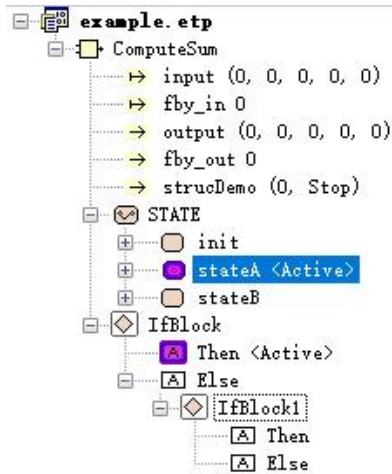

Figure 39. SCADE Suite simulation

| | |
|---|---|
| **Operation Details: ComputeSum** | |
| Name | Value |
| ▼ Parameters | |
| input | {(0\|->0),(1\|->0),(2\|->0),(3\|->0),(4\|->0)} |
| fby_in | 0 |
| ▼ Return values | |
| output | {(0\|->0),(1\|->0),(2\|->0),(3\|->0),(4\|->0)} |
| fby_out | 0 |
| strucDemo | rec(fby_data:0,move:Stop) |

Figure 40. ProB simulation

| Name | Value | Previous Value |
|---|---|---|
| ▼ VARIABLES | | |
| store | {(0↦0),(1↦0),(2↦0)} | {(0↦0),(1↦0),(2↦0)} |
| sm_state | stateA | init |

Figure 41. ProB simulation

From the figures above, it can be observed that the outputs (*output, fby_out, and strucDemo*) of both models are completely identical. Additionally, the state machine transitions from *init* to *stateA*.

B. In the second cycle, we set the input values as *input = [1, 2, 3, 4, 5]* and *fby_in = 1*. The step-by-step execution results are shown in the figures below, where Figure 42 displays the simulation result of the SCADE model, Figure 43 and Figure 44 show the simulation result of the B-Model.

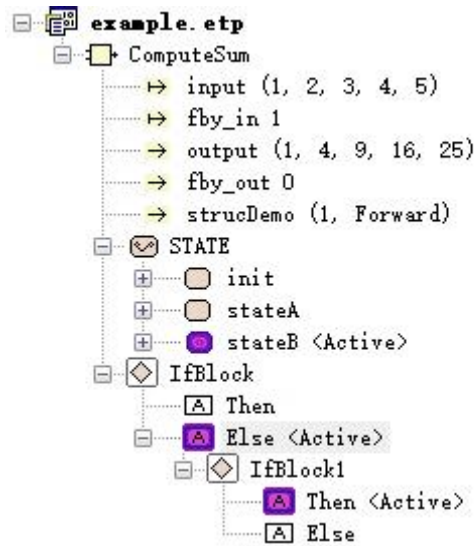

Figure 42. SCADE Suite simulation

| | |
|---|---|
| **Operation Details: ComputeSum** | |
| Name | Value |
| ▼ Parameters | |
| input | {(0\|->1),(1\|->2),(2\|->3),(3\|->4),(4\|->5)} |
| fby_in | 1 |
| ▼ Return values | |
| output | {(0\|->1),(1\|->4),(2\|->9),(3\|->16),(4\|->25)} |
| fby_out | 0 |
| strucDemo | rec(fby_data:1,move:Forward) |

Figure 43. ProB simulation

Figure 44. ProB simulation

From the figures above, it can be observed that the outputs (*output, fby_out, and strucDemo*) of both models remain completely identical. Additionally, the state machine transitions from *stateA* to *stateB*.

C.  In the third cycle, we set the input values as *input = [6, 7, 8, 9, 10]* and *fby_in = 2*, with the step-by-step execution results shown in the figures below, where Figure 45 displays the simulation result of the SCADE model, Figure 46 and Figure 47 shows the simulation result of the B-Model.

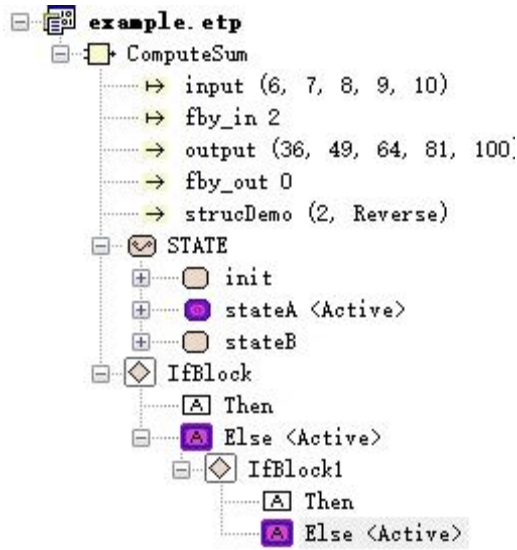

Figure 45. SCADE Suite simulation

Figure 46. ProB simulation

Figure 47. ProB simulation

From the figures above, it can be observed that the outputs (*output*, *fby_out*, and *strucDemo*) of both models remain completely identical. Additionally, the state machine transitions from *stateB* back to *stateA*.

D. In the fourth cycle, we set the input values as *input = [6, 7, 8, 9, 10]* and *fby_in = 3*, with the step-by-step execution results displayed in the figures below, where Figure 48 shows the simulation result of the SCADE model, Figure 49 and Figure 50 presents the simulation result of the B-Model.

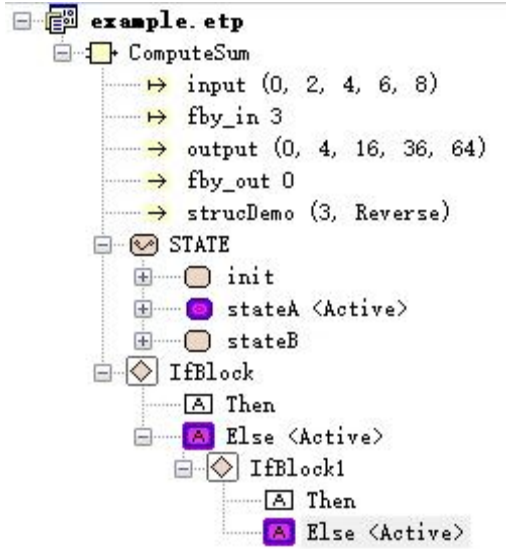

Figure 48. SCADE Suite simulation

| Name | Value |
|---|---|
| ▼ Parameters | |
| input | {(0|->0),(1|->2),(2|->4),(3|->6),(4|->8)} |
| fby_in | 3 |
| ▼ Return values | |
| output | {(0|->0),(1|->4),(2|->16),(3|->36),(4|->64)} |
| fby_out | 0 |
| strucDemo | rec(fby_data:3,move:Reverse) |

Figure 49. ProB simulation

| Name | Value | Previous Value |
|---|---|---|
| ▼ VARIABLES | | |
| store | {(0↦1),(1↦2),(2↦3)} | {(0↦0),(1↦1),(2↦2)} |
| sm_state | stateA | stateA |

Figure 50. ProB simulation

From the figures above, it can be observed that the outputs (*output, fby_out,* and *strucDemo*) of both models remain completely identical. Additionally, the state machine stays in *stateA*.

E. In the fifth cycle, we set the input values as *input = [1, 3, 5, 7, 9]* and *fby_in = 4*, with the step-by-step execution results displayed in the figures below, where Figure 51 shows the simulation result of the SCADE model, Figure 52 and Figure 53 presents the simulation result of the B-Model.

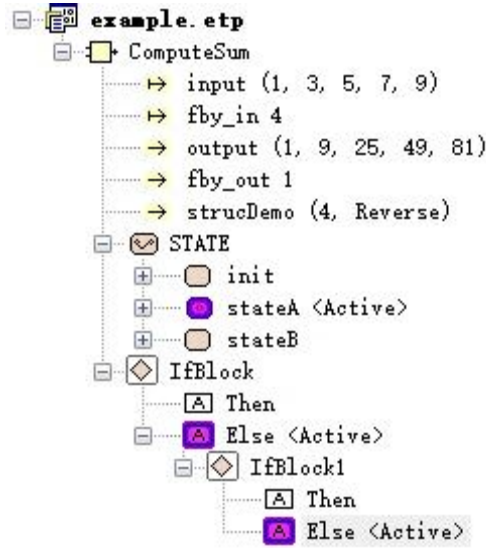

Figure 51. SCADE Suite simulation

Figure 52. ProB simulation

Figure 53. ProB simulation

From the figures above, it can be observed that the outputs (*output, fby_out,* and *strucDemo*) of both models remain completely identical. Additionally, the state machine stays in *stateA*, and the *fby_in* values from the last three cycles have been correctly stored.

The experimental results demonstrate that the translation from the SCADE model to the B-Model is correct, as both models exhibit identical behavior under the same inputs.

## 5. Experiment 2

This experiment demonstrates how translating a SCADE model into a B-Model allows for formal verification using the B-Model's powerful descriptive capabilities. The experiment verifies whether the SCADE model meets specific safety requirements, and through counterexample detection and model refinement, ensures that the model fully complies with safety requirements. This method provides a systematic process that can be applied to the design and verification of other safety-critical systems. The greatest advantage of translating SCADE into a B-Model is that it enables the use of set theory, relational algebra, and first-order predicate logic to express safety requirements, significantly enhancing its descriptive power compared to SCADE. Below, we present an example that demonstrates how formal verification of the B-Model can be used to achieve formal verification of the SCADE model.

The example is a simplified communication protocol consisting of two state machines: *CON_STATE*, which determines different communication states based on the input, and *PRO_STATE*, which determines whether data processing can be performed based on the communication state (with the actual data processing mechanism omitted in this study). The SCADE model, shown in Figure 54 and Figure 55, represents these two state machines, where the input *input_event* is an enumeration type defined in Figure 56, and the output *process_enable* is a flag indicating whether data processing is allowed. The text-based SCADE model is provided in Appendix 3, and the translated B-Model is provided in Appendix 4.

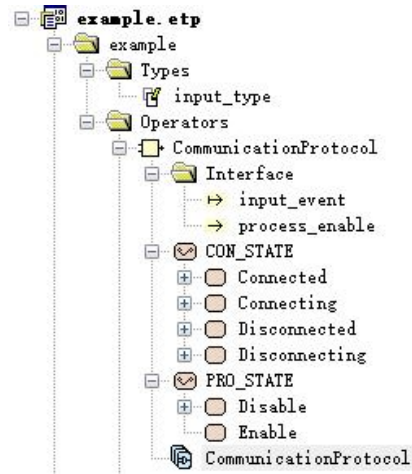

Figure 54. SCADE Suite model structure

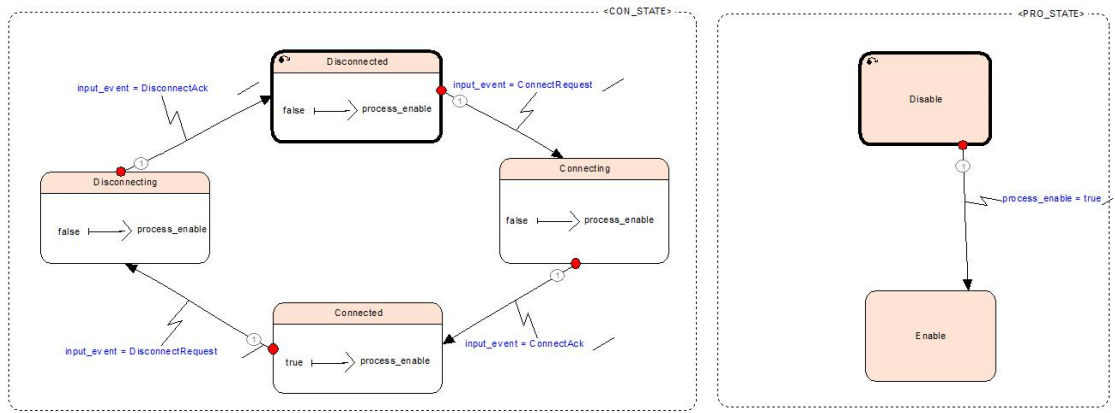

Figure 55. SCADE Suite state machine

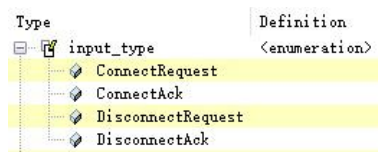

Figure 56. SCADE Suite type definitions

A safety requirement for this model specifies that data processing is only allowed in the Connected state. To verify whether the SCADE model satisfies this requirement, an invariant condition is added under the *INVARIANT* section of the corresponding B-Model.

$$(process\_state = Enable \Rightarrow connection\_state = Connected) \land$$
$$(connection\_state \neq Connected \Rightarrow process\_state = Disable)$$

The meaning of this invariant is that when *process_state* is in the *Enable* state, it implies that *connection_state* is in the *Connected* state. Similarly, if *connection_state* is not in the *Connected* state, it implies that *process_state* is in the *Disable* state. This invariant serves as a formal description of the requirement that data processing is only allowed in the *Connected* state.

Start ProB, load the B model, and select Invariant in the verifier. Additionally, a symbolic model checking approach, such as the classic K-induction algorithm, can be applied. Run the verifier, and the results are shown in Figure 57.

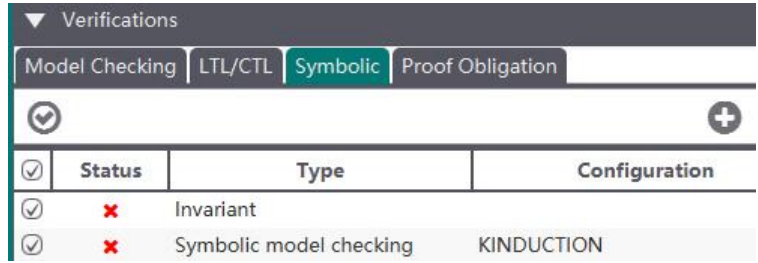

Figure 57. ProB Formal Verification

After running the verification, we observe that it fails, and by examining the counterexample, we find that after a sequence of inputs (*ConnectRequest* → *ConnectAck* → *DisconnectRequest*), the safety requirement is violated, as shown in Figure 58; at this point, *connection_state* is in *Disconnecting*, but *process_state* remains *Enable*, contradicting the safety requirement, thereby proving that the model does not satisfy the safety requirement, as shown in Figure 59.

| Position▲ | Transition |
|---|---|
| 0 | ---root--- |
| 1 | INITIALISATION |
| 2 | HandleEvent(input_event=ConnectRequest) → FALSE |
| 3 | HandleEvent(input_event=ConnectAck) → TRUE |
| 4 | HandleEvent(input_event=DisconnectRequest) → FALSE |

Figure 58. ProB Formal Verification

| Name | Value | Previous Value |
|---|---|---|
| ▼ VARIABLES | | |
| connection_state | Disconnecting | Connected |
| process_state | Enable | Enable |
| CONSTANTS | | |
| ▶ SETS | | |
| ▼ INVARIANT | false | true |
| ▶ [⇒] process_state = Enable ⇒ connection_state = Connected | false | true |
| ▶ [⇒] connection_state ≠ Connected ⇒ process_state = Disable | false | true |

Figure 59. ProB Formal Verification

After analysis, as shown in the figure, the *PRO_STATE* state machine is missing a transition from *Enable* to *Disable*; after modifying the SCADE model, the updated version is shown in Figure 60, with the text-based SCADE model available in Appendix 5 and the corresponding B-Model in Appendix 6.

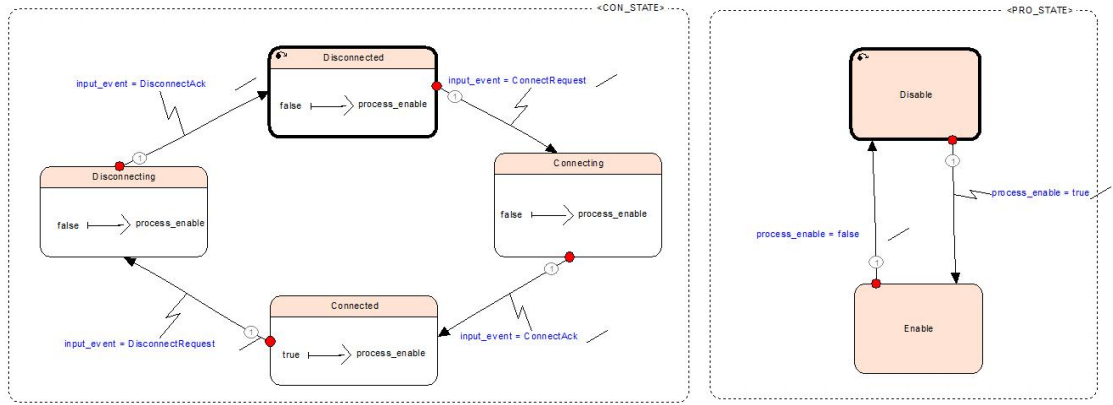

Figure 60. SCADE Suite state machine

After loading the B-Model into ProB and running the verifier, the results are shown in Figure 61. At this point, the verification confirms that the model fully satisfies the safety requirement.

Figure 61. ProB formal verification

## 6. Conclusion

This paper proposes a method for transforming SCADE models into B models, leveraging the B-Method's powerful specification capabilities and mature formal verification tools to compensate for SCADE's limitations in abstract specification verification. Through experimental validation, we have demonstrated the effectiveness and correctness of this method, ensuring that the logical consistency of the SCADE model is preserved after transformation into a B model.

We have introduced specific transformation rules that cover various components, including data types, constants, equations, conditional expressions, sequential operators, higher-order operators, and state machines. These rules ensure semantic and functional consistency between the SCADE model and the transformed B model. Through experiments, we have validated the correctness of these transformation rules, confirming that the SCADE and B models produce identical simulation results under the same input conditions.

Furthermore, we demonstrated how to use B-Model formal verification tools (such as ProB) to formally verify SCADE models. By converting SCADE models into B models, we can formally express safety requirements within the B model and use B-Method verification tools to check whether these requirements are satisfied. The experimental results indicate that this approach effectively identifies potential issues within SCADE models and ensures that the model meets design specifications.

In summary, the proposed method not only enhances the formal verification capabilities of SCADE models but also provides a systematic process for the design and verification of safety-critical systems. By transforming SCADE models into B models, we can leverage the B-Method's powerful descriptive capabilities and advanced verification tools, thereby improving the overall verification capabilities of safety-critical systems and providing additional support for model analysis.

Future research can explore more complex transformation rules for SCADE models and investigate their validation and optimization in broader practical applications.

## 7. References

[1] Colaço J L, Pagano B, Pouzet M. SCADE 6:A formal language for embedded critical software development. 2017 International Symposium on Theoretical Aspects of Software Engineering(TASE). IEEE,2017:1-11.

[2] Ferrari, Alessio, and Maurice H. Ter Beek. Formal methods in railways: a systematic mapping study. ACM Computing Surveys, 2022, 55, 4: 1-37.

[3] Abrial, J. R. The B-Book: Assigning Programs to Meanings. Cambridge: Cambridge University Press, 1996.

[4] Lecomte, T., Deharbe, D., Prun, E., Mottin, E. Applying a formal method in industry: a 25-year trajectory. In: Cavalheiro, S., Fiadeiro, J. (eds.) SBMF (2017). LNCS, vol. 10623, 70–87. Cham: Springer.

[5] Jahier, E., Raymond, P. and Halbwachs, N.. The Lustre V6 reference manual. Verimag, Grenoble, Dec, 2016.

[6] Cansell, D., Méry, D.. Foundations of the B method. Computing And Informatics, 2003, 22(3/4), 221-256.

[7] Boulanger, J. L. (Ed.). Formal Methods Applied to Complex Systems: Implementation of the B Method. John Wiley & Sons, 2014.

[8] Hoang, T. S. An introduction to the Event-B modeling method. Industrial Deployment of System Engineering Methods, 2013, 211-236.

[9] ClearSy. Atelier B, User Manual Version 4.0. Retrieved from http://www.atelierb.eu/2021.

[10] Leuschel, M., Butler, M.J. ProB: an automated analysis toolset for the B method. STTT 10(2), 2008, 185–203.

[11] ANSYS. SCADE Suite, Scade Language Reference Manual, Published May 2020. Retrieved from https://developer.ansys.com/docs/scade.

# Appendix 1: Text-Based SCADE Model for Experiment 1

```
automaton STATE
    initial state init
    unless
        if fby_in = 0
            restart stateA;

    state stateA
    unless
        if fby_in = 1
            restart stateB;

    state stateB
    unless
        if fby_in > 1
            restart stateA;

returns .. ;
output = L12;
L13 = input;
L12 = (map $*$ <<MAX_SIZE>>)(L13, L14);
L11 = fby(L10; 3; 0);
L10 = fby_in;
fby_out = L11;
activate IfBlock
    if (fby_in = 0) then
        var
            L1 : strucType;
            L2 : uint8;
            L3 : MOVE;
        let
            L1 = (make strucType)(L2, L3);
            strucDemo = L1;
            L2 = fby_in;
            L3 = Stop;
        tel
    else
        activate IfBlock1
            if (fby_in = 1) then
                var
                    L4 : uint8;
                    L5 : strucType;
                    L6 : MOVE;
                let
                    strucDemo = L5;
                    L5 = (make strucType)(L4, L6);
                    L4 = fby_in;
                    L6 = Forward;
                tel
            else
                var
                    L7 : strucType;
                    L8 : uint8;
                    L9 : MOVE;
                let
                    strucDemo = L7;
                    L7 = (make strucType)(L8, L9);
                    L8 = fby_in;
                    L9 = Reverse;
                tel

        returns .. ;
```

# Appendix 2: B-Model for Experiment 1

```
MACHINE
    example

SETS
    // Enumerated types, equivalent to enums in SCADE
    MOVE = {Stop, Forward, Reverse};
    // States for the state machine
    STATE = {init, stateA, stateB}

CONSTANTS
    // Basic data types and constant
    uint8_t, MAX_SIZE

PROPERTIES
    // Define the value ranges for basic types and the value of constant
    uint8_t = 0..255 ∧
    MAX_SIZE = 5

VARIABLES
    // Sequential Operator declarations
    store,
    // State Variable declaration
    sm_state

INVARIANT
    // Variable type definitions
    sm_state ∈ STATE ∧
    store ∈ 0..2 → uint8_t

INITIALISATION
    // Initialize variables
    sm_state := init ||
    store := { 0 ↦ 0, 1 ↦ 0, 2 ↦ 0 }

OPERATIONS
    // Contains higher-order operators, sequential operators, and state machines
    output, fby_out, strucDemo ← ComputeSum(input, fby_in) =
        PRE
            input ∈ 0..(MAX_SIZE - 1) → uint8_t ∧
            fby_in ∈ uint8_t
        THEN
            output := {0 ↦ 0, 1 ↦ 0, 2 ↦ 0, 3 ↦ 0, 4 ↦ 0};
            fby_out := 0;
            strucDemo := rec(fby_data ∈ 0, move ∈ Stop);

            // Higher-order operator using WHILE loop to simulate map
            VAR idx IN
                idx := 0;
                WHILE idx < MAX_SIZE DO
                    output(idx) := input(idx) * input(idx);
                    idx := idx + 1
                INVARIANT
                    ∀i.(i ∈ 0..(idx - 1) ⇒ output(i) = input(i) * input(i))
                VARIANT
                    MAX_SIZE - idx
                END
            END;

            // Sequential operator simulate fby
            fby_out := store(0);
            store(0) := store(1);
            store(1) := store(2);
            store(2) := fby_in;

            // State machine
            CASE sm_state OF
                EITHER init THEN
                    IF fby_in = 0 THEN
                        sm_state := stateA
                    END
                OR stateA THEN
                    IF fby_in = 1 THEN
                        sm_state := stateB
                    END
                OR stateB THEN
                    IF fby_in > 1 THEN
                        sm_state := stateA
                    END
                END
            END;

            // Example usage of IF condition. enums and structure
            IF fby_in = 0 THEN
                strucDemo := rec(fby_data ∈ fby_in, move ∈ Stop)
            ELSE
                IF fby_in = 1 THEN
                    strucDemo := rec(fby_data ∈ fby_in, move ∈ Forward)
                ELSE
                    strucDemo := rec(fby_data ∈ fby_in, move ∈ Reverse)
                END
            END
        END
END
```

**Appendix 3: Text-Based SCADE Model for Experiment 2 (Version 1)**

```
automaton CON_STATE
    initial state Disconnected
    unless
        if input_event = ConnectRequest
            restart Connecting;

    var
        L1 : bool;
    let
        L1 = false;
        process_enable = L1;
    tel
    state Connecting
    unless
        if input_event = ConnectAck
            restart Connected;

    var
        L2 : bool;
    let
        process_enable = L2;
        L2 = false;
    tel
    state Connected
    unless
        if input_event = DisconnectRequest
            restart Disconnecting;

    var
        L3 : bool;
    let
        process_enable = L3;
        L3 = true;
    tel
    state Disconnecting
    unless
        if input_event = DisconnectAck
            restart Disconnected;

    var
        L4 : bool;
    let
        process_enable = L4;
        L4 = false;
    tel
returns .. ;
automaton PRO_STATE
    initial state Disable
    unless
        if process_enable = true
            restart Enable;

    state Enable
returns .. ;
```

# Appendix 4: B-Model for Experiment 2(Version 1)

```
MACHINE CommunicationProtocol
SETS
    // Define the set of states for the protocol.
    CON_STATE = {Disconnected, Connecting, Connected, Disconnecting};
        PRO_STATE = {Enable, Disable};
    // Define the set of input events.
    INPUT_EVENT = {ConnectRequest, ConnectAck, DisconnectRequest, DisconnectAck}

VARIABLES
    connection_state, process_state

INVARIANT
    // Variable type constraints.
    connection_state ∈ CON_STATE ∧
        process_state ∈ PRO_STATE ∧
        // Safety requirement: Data can only be processed in the Connected state.
        (process_state = Enable ⇒ connection_state = Connected) ∧
    (connection_state ≠ Connected ⇒ process_state = Disable)

INITIALISATION
    // Initialize the connection state to Disconnected.
    connection_state := Disconnected ||
        // Initialize the data processing state to Disabled.
        process_state := Disable

OPERATIONS
    // Process input events and perform state transitions.
    process_enable ← HandleEvent(input_event) =
        PRE
            input_event ∈ INPUT_EVENT
        THEN
                        CASE connection_state OF
                                EITHER Disconnected THEN
                                        IF input_event = ConnectRequest THEN
                                                connection_state := Connecting;
                                                process_enable := FALSE
                                        ELSE
                                                process_enable := FALSE
                                        END
                                OR Connecting THEN
                                        IF input_event = ConnectAck THEN
                                                connection_state := Connected;
                                                process_enable := TRUE
                                        ELSE
                                                process_enable := FALSE
                                        END
                                OR Connected THEN
                                        IF input_event = DisconnectRequest THEN
                                                connection_state := Disconnecting;
                                                process_enable := FALSE
                                        ELSE
                                                process_enable := TRUE
                                        END
                                OR Disconnecting THEN
                                        IF input_event = DisconnectAck THEN
                                                connection_state := Disconnected;
                                                process_enable := FALSE
                                        ELSE
                                                process_enable := FALSE
                                        END
                                END
                        END;

                        CASE process_state OF
                                EITHER Disable THEN
                                        IF process_enable = TRUE THEN
                                                process_state := Enable
                                        END
                                OR Enable THEN
                                        skip
                                END
                        END
            END
        END
```

**Appendix 5: Text-Based SCADE Model for Experiment 2 (Version 2)**

```
automaton CON_STATE
    initial state Disconnected
    unless
        if input_event = ConnectRequest
            restart Connecting;

    var
        L1 : bool;
    let
        L1 = false;
        process_enable = L1;
    tel
    state Connecting
    unless
        if input_event = ConnectAck
            restart Connected;

    var
        L2 : bool;
    let
        process_enable = L2;
        L2 = false;
    tel
    state Connected
    unless
        if input_event = DisconnectRequest
            restart Disconnecting;

    var
        L3 : bool;
    let
        process_enable = L3;
        L3 = true;
    tel
    state Disconnecting
    unless
        if input_event = DisconnectAck
            restart Disconnected;

    var
        L4 : bool;
    let
        process_enable = L4;
        L4 = false;
    tel
returns .. ;
automaton PRO_STATE
    initial state Disable
    unless
        if process_enable = true
            restart Enable;

    state Enable
    unless
        if process_enable = false
            restart Disable;

returns .. ;
```

## Appendix 6: B-Model for Experiment 2(Version 2)

```
MACHINE CommunicationProtocol
SETS
    // Define the set of states for the protocol.
    CON_STATE = {Disconnected, Connecting, Connected, Disconnecting};
        PRO_STATE = {Enable, Disable};
    // Define the set of input events.
    INPUT_EVENT = {ConnectRequest, ConnectAck, DisconnectRequest, DisconnectAck}

VARIABLES
    connection_state, process_state

INVARIANT
    // Variable type constraints.
    connection_state ∈ CON_STATE ∧
        process_state ∈ PRO_STATE ∧
        // Safety requirement: Data can only be processed in the Connected state.
        (process_state = Enable ⇒ connection_state = Connected) ∧
    (connection_state ≠ Connected ⇒ process_state = Disable)

INITIALISATION
    // Initialize the connection state to Disconnected.
    connection_state := Disconnected ||
        // Initialize the data processing state to Disabled.
        process_state := Disable

OPERATIONS
    // Process input events and perform state transitions.
    process_enable ⟵ HandleEvent(input_event) =
        PRE
            input_event ∈ INPUT_EVENT
        THEN
                        CASE connection_state OF
                            EITHER Disconnected THEN
                                    IF input_event = ConnectRequest THEN
                                            connection_state := Connecting;
                                            process_enable := FALSE
                                    ELSE
                                            process_enable := FALSE
                                    END
                            OR Connecting THEN
                                    IF input_event = ConnectAck THEN
                                            connection_state := Connected;
                                            process_enable := TRUE
                                    ELSE
                                            process_enable := FALSE
                                    END
                            OR Connected THEN
                                    IF input_event = DisconnectRequest THEN
                                            connection_state := Disconnecting;
                                            process_enable := FALSE
                                    ELSE
                                            process_enable := TRUE
                                    END
                            OR Disconnecting THEN
                                    IF input_event = DisconnectAck THEN
                                            connection_state := Disconnected;
                                            process_enable := FALSE
                                    ELSE
                                            process_enable := FALSE
                                    END
                            END
                    END;

                        CASE process_state OF
                            EITHER Disable THEN
                                    IF process_enable = TRUE THEN
                                            process_state := Enable
                                    END
                            OR Enable THEN
                                    IF process_enable = FALSE THEN
                                            process_state := Disable
                                    END
                            END
                    END
            END
    END
```